# ESTIMATING REQUIRED FLEXIBILITY FOR SECURE DISTRIBUTION GRID OPERATION CONSIDERING THE UNCERTAINTIES OF EV AND PV

Manijeh ALIPOUR
Depsys, Puidoux - Switzerland
manijeh.alipour@depsys.com

Omid ALIZADEH-MOUSAVI
Depsys, Puidoux - Switzerland
omid.mousavi@depsys.com

## ABSTRACT

*Renewable energy productions and electrification of mobility are promising solutions to reduce greenhouse gas emissions. Their effective integration in a power grid encounters several challenges. The uncertain nature of renewable energy productions as well as stochastic consumptions of electric vehicles introduce remarkable intermittency to a distribution grid and results in bi-uncertain characteristics of both supply and demand sides. One way to verify the secure grid operation within acceptable voltage and loading levels is to assess its required flexibility considering possible boundaries of uncertain variables. In this paper, first a comprehensive linear model of distribution grid considering all pertaining constraints is presented. Then, a flexibility estimation technique is proposed based on the feasibility study of the uncertain space of photovoltaic power productions and load containing electric vehicles. The proposed methodology uses grid monitoring data to determine grid state and to model uncertain parameters. The model is applied on a real low voltage (LV) system equipped with grid monitoring devices.*

## INTRODUCTION

The deployment of renewable energy productions has many advantages, comprising economic convenience, reducing the reliance on fossil fuel markets (especially, gas and oil) and environmental friendliness. In addition, the wide penetration of renewable energy sources accelerates occupation in the EU, by job creation in different 'green' technologies. Aim behind the European Green Deal (COM(2019) 640 final) is to be the world's first climate-neutral continent by 2050. In spite of renewable generation's significant advantages, it has the defects of uncertainty and fluctuation. Besides to renewable generations, electric vehicles as a new variable load increase the intermittency of distribution grid and leads to bi-uncertain characteristics of both demand and supply sides. With the increased penetration of these uncertainty sources, the modern power system is confronted with a challenge to preserve the reliability, security, and quality of supply. Thus, a more flexible distribution grid is required.

The distribution grid flexibility can be defined from different perspectives [1]. In [2], the flexibility is specified as the degree to which a system can change its electricity consumption and production in response to anticipated or unanticipated variabilities. The flexibility envelope method is presented in [3] to assess the flexibility potential of individual power system assets and their aggregation at the system level. A flexibility measure is developed in [4] which indicates the largest uncertainty deviation that a system can bear. In [5] an algorithm is proposed which allocates aggregate-level control decisions amongst individual systems economically. Concerning the flexibility estimation of distribution grids, an index is proposed in [6] that is related to specific viewpoints such as power regulation ability and energy balance ability of distribution grids. In [7] the net load uncertainty is considered in the flexibility assessment of distribution grids.

The secure grid operation can be verified by assessing the required flexibility considering boundaries of uncertain variables. A hyper-rectangle is implemented for heat exchanger networks [8] to define the multi-dimensional region of uncertain variables. In this paper, the flexibility and the lack of flexibility are quantified by determining acceptable boundaries of uncertain parameters in distribution grids.

A large portion of uncertainty resources are not monitored by the Distribution System Operator (DSO) and as a consequence, the level of available flexibility of the grid as well as the lack of flexibility caused by the uncertain resources are not identified. In such a situation, the power flows from unmonitored and uncontrolled resources may cause voltage violations and congestions. Therefore, the monitoring of grid and the control of resources connected to the grid are vital for the secure operation of distribution grid.

The DSOs around the world are going through the roll-out of smart metering and grid monitoring devices [9, 10]. However, equipping all the nodes of distribution grid with monitoring devices is practically impossible due to costs of required underlying infrastructure. In addition, although all the grid topology information is usually available in the DSO's Geographic Information System (GIS), a trustable and up-to-date data of LV grid topology and parameters is not available to take decisions for the secure grid operation [11]. To face these challenges, the grid sensitivity coefficients of a subset of grid nodes can be used as an approximation of the power flow model. The sensitivity coefficients can be calculated between a plurality of measurement nodes without using the information of grid parameters [13], hereafter called model-less approach. This approach significantly reduces the required number of monitored nodes, thus reducing the cost of required grid monitoring infrastructure, and does not rely on the availability of an accurate and up-to-date grid parameters. This paper proposes a flexibility estimation model for distribution grids. The main contributions are as follows:

1) The proposed flexibility estimation method is based on the feasibility study of the uncertain space of load containing electric vehicles and photovoltaic powers. It allows evaluating the flexibility and lack of flexibility based on the coverage of the feasible space to the

2) The projection of each direction in the uncertain space to the feasible space is illustrated and investigated. The illustrated projection to the feasible space can assist in determining the reason of flexibility inadequacy.
3) The linear power flow model based on the sensitivity coefficients is used to model the grid operation constraints. It properly accounts for the secure grid operation within acceptable voltage levels and without congestions.
4) The proposed model for the calculation of the flexibility index is a linear and convex mathematical optimization problem that can be solved efficiently with non-commercial solvers and the optimality of the solution is guaranteed.

The remaining parts of the paper is structured as following. Next section provides the definition of distribution grid flexibility. Then, the proposed mathematical formulation is presented. The results of applying the method on a real LV grid are evaluated. Finally, the conclusions and outcomes are discussed.

## THE FLEXIBILITY OF DISTRIBUTION GRID

In this paper, the flexibility of distribution grid is described as the capability of distribution grid to effectively cope with multiple uncertainties of grid operation. Fig. 1 illustrates the structure of the proposed flexibility estimation algorithm in a LV distribution grid. The power grid contains Electric Vehicles (EV), EV charging stations, photovoltaic arrays and commercial load. In the proposed method, grid monitoring data are used to calculate the grid state and the sensitivity coefficients as well as to model uncertain parameters. Fig. 1 shows the way that flexibility quantification is realized using the grid monitoring data.

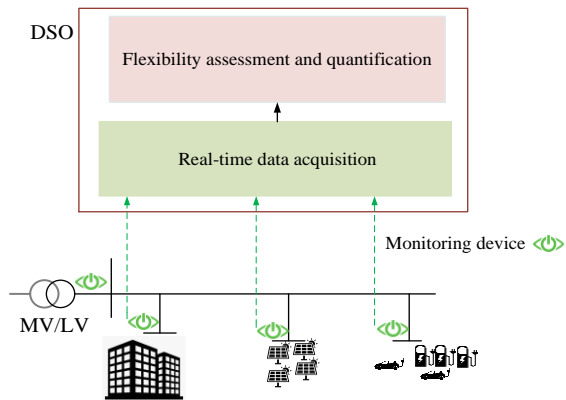

Fig. 1. Proposed monitoring based flexibility quantification structure

In fact, the proposed model is accomplished by detailed and accurate information of the resources from the grid monitoring facilities. Fig. 2 shows the concept of proposed flexibility quantification method. The uncertain space is calculated using historic data and prediction methods. The margin of feasible space characterizes the maximum feasible deviation from the expected operation point 'O'. Any operation point in the uncertain space can be characterized by two components: the direction vector and the normalized variation value. The direction matrix expresses the direction vector, in which each diagonal element indicates an uncertain variable, and the range of element's value is [−1, 1] [12]. In order to guarantee that direction matrix elements represent a unique direction, at least one of the direction matrix elements must be set to -1 or 1. This will refrain the duplication of direction matrices with the same direction like *diag(1, 0.5)* and *diag(0.8, 0.4)*.

Any point outside the feasible region indicates an infeasible grid operation point where the technical constraints of the grid are not satisfied. The minimum feasible deviation direction stands for the critical direction. The boundary point, shown by 'C' in Fig.2, corresponding to the critical direction is the flexibility index which reveals the adequacy or insufficiency of flexibility resources in the system.

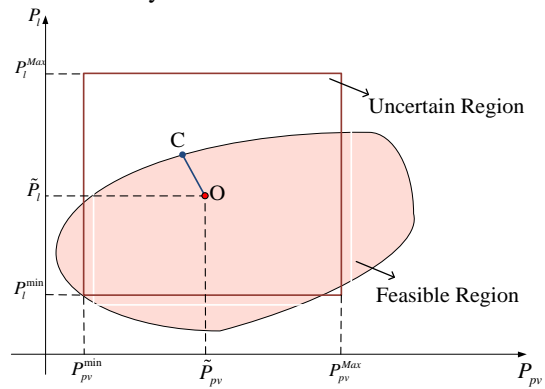

Fig. 2. The concept of the proposed flexibility quantification method

## MATHEMATICAL FORMULATION

A mathematical optimization problem is formulated to determine the feasible space. The objective function is the minimum value of maximum feasible variation in the direction D ($\beta^D$) of feasible operation region. In addition, the flexibility (F) is defined as the minimum value of $\beta^D$.

$$F = min\beta^D \qquad (1)$$
$$\beta^D = max(\beta) \qquad (2)$$
$$X = \tilde{X} + \beta D \Delta X \qquad (3)$$

where $X$ and $\tilde{X}$ represent the uncertain parameter and its expected value which are distribution grid's load and PV generation. $\Delta X$ indicates the expected value and maximum/minimum value of uncertain parameter's difference. If the flexibility index is equal to one (F=1), the flexibility resources are enough for the secure operation of the grid. In addition, it means that feasible region covers the uncertain region. However, if the flexibility index is less than one (F <1), there is not sufficient margin for the technically secure grid operation and additional flexibility resources are required. The direction matrix $D$ is a diagonal matrix that can be defined as:

$$D = \begin{bmatrix} d_1 & & & \\ & d_2 & & \\ & & \ddots & \\ & & & d_n \end{bmatrix} \qquad (4)$$

where diagonal element $d_i$ is the direction on the $i^{th}$ uncertain parameter and $n$ is the number of uncertain parameters.

For the flexibility estimation a linear power flow model is used based on the sensitivity coefficients calculated using the model-less approach. The formulation is described below.

$$\Delta V_m = \sum_n [K_{m,n}^{VP} \Delta P_n + K_{m,n}^{VQ} \Delta Q_n] \quad (5)$$
$$\Delta I_l = \sum_n [K_{l,n}^{IP} \Delta P_n + K_{l,n}^{IQ} \Delta Q_n] \quad (6)$$
$$V^{min} \leq V_m^0 + \Delta V_m \leq V^{max} \quad (7)$$
$$-I^{max} \leq I_l^0 + \Delta I_l \leq I^{max} \quad (8)$$
$$\Delta V_m = V_m - V_m^0 \quad (9)$$
$$\Delta I_l = I_l - I_l^0 \quad (10)$$
$$0 \leq S^{grid} \leq s_{Trafo}^{max} \quad (11)$$

The voltage and current sensitivity coefficients are used in equations (5) and (6), respectively, to model the impacts of nodal active and reactive power changes ($\Delta P_n$ and $\Delta Q_n$) on the nodal voltage variations ($\Delta V_m$) and branch current variations ($\Delta I_l$). The sensitivity coefficients, i.e., $K^{VP}, K^{VQ}, K^{IP}$ and $K^{IQ}$, are computed around the grid operation point corresponding to voltage $V_m^0$ and current $I_{l,t}^0$. The voltage level at each node and the branch current should be within the allowed limits as given in (7) and (8), respectively. The voltage and current deviations can be calculated as given by equations (9) and (10). The constraint (11) limits the apparent power flow in the MV/LV transformer ($S^{grid}$) by the capacity of transformer ($s_{max}^{Trafo}$).

## RESULTS AND DISCUSSIONS

In this section, the performance of proposed flexibility calculation method is tested and validated in a modified LV grid of Switzerland. The LV grid includes a PV unit with known capacity to the DSO, and connected to bus 101 with the capacity of 72 kVA. Further, EVSEs are connected to the buses 102 and 106. The grid with several GridEye monitoring devices is illustrated in Fig. 3. The real-time information of the aggregated loads and productions are provided by the GridEye monitoring devices. Figs. 4 and 5 illustrate the 10-minute granularity of PV generation and load profiles used in the simulations. In this work, the DSO's objective is to estimate the distribution grid's flexibility value, the value of flexibility insufficiency and the periods with insufficient flexibility.

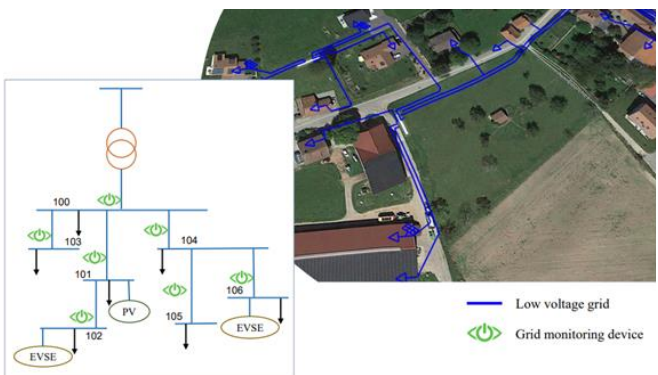

Fig. 3. The LV grid topology with grid monitoring devices

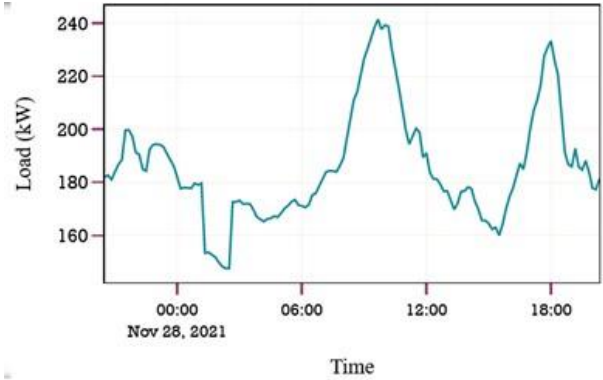

Fig. 4. Load profile of LV grid

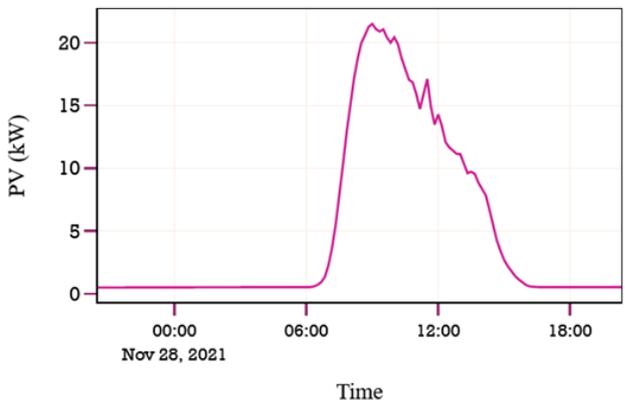

Fig. 5. PV generation at node 101

Table 1 provides the simulation results for time slots that there is insufficient flexibility in the grid, considering the uncertainties of load and PVs. For the remaining time slots, the flexibility value is equal to one indicating that there is adequate flexibility in the grid. At the time periods from 09:20 to 10:10 and from 17:40 to 18:20 on Nov 28, 2021 due to the high level of load and the thermal capacity of lines, the flexibility is less than one. The uncertain and feasible regions for two sample time slots are depicted in Figs. 6 and 7. The value of flexibility at time slot 18:00 (Fig. 7) has the lowest value 0.49. In this time slot, the load level is higher than the average consumption and the PV generation has the lowest value. Moreover, at time slot 09:40 (Fig. 6), the load is at its highest value and the value of flexibility is 0.74. At this time slot, although the load has the highest value, the flexibility index is higher than the one at time slot 18:00. The higher level of flexibility is due to the high level of PV generation at this time which is 20.4 kW.

Table 1. Flexibility results for insecure time slots

| Time slot | Flexibility |
|---|---|
| Nov 28, 2021, 09:20 | 0.93 |
| Nov 28, 2021, 09:30 | 0.84 |
| Nov 28, 2021, 09:40 | 0.74 |
| Nov 28, 2021, 09:50 | 0.84 |
| Nov 28, 2021, 10:00 | 0.80 |
| Nov 28, 2021, 10:10 | 0.80 |
| Nov 28, 2021, 17:40 | 0.67 |
| Nov 28, 2021, 17:50 | 0.57 |
| Nov 28, 2021, 18:00 | 0.49 |
| Nov 28, 2021, 18:10 | 0.72 |
| Nov 28, 2021, 18:20 | 0.90 |

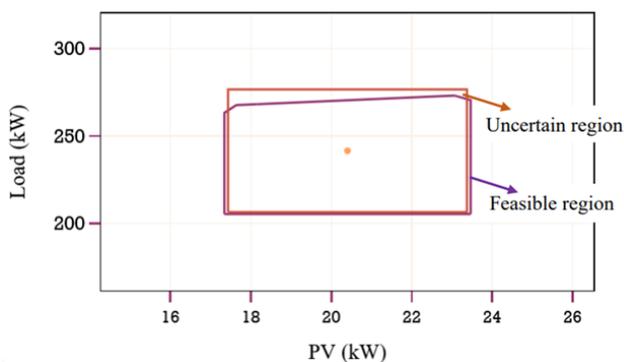

Fig. 6. Feasible and uncertain regions on Nov 28, 2021 at 09:40.

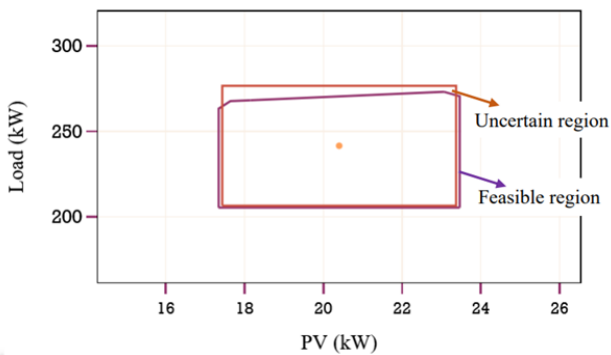

Fig. 7. Feasible and uncertain regions on Nov 28, 2021 at 18:00.

## CONCLUSIONS

In this paper, a flexibility estimation methodology is proposed taking into account the feasible grid operation region and the uncertain region of load containing electric vehicle and photovoltaic powers. The method is applicable by using the information of grid monitoring devices. The simulation results on the LV grid illustrated the inadequate flexibility in time slots with the higher level of load and the lower level of PV generation than the average level. However, the model can detect other factors like voltage limit violation regarding the grid constraints that limit the grid's flexibility. The outcome of proposed model informs the grid operator regarding the time slots with sufficient and insufficient flexibility along with the value of insufficiency considering the uncertainty space.

## Acknowledgments

This project is supported by the European Union's Horizon 2020 programme under the Marie Sklodowska-Curie grant agreement no. 101026259.